\documentclass[a4paper,14pt]{article}
\usepackage[font={small}]{caption}
\usepackage[english]{babel}
\usepackage{graphicx}
\usepackage{caption}
\usepackage{indentfirst}
\usepackage{epsfig}
\usepackage{authblk}
\usepackage{geometry}
\usepackage{natbib}
\usepackage{wrapfig}
\usepackage{lipsum}

\title{The Influence of Coronal Mass Ejections on the Gas Dynamics of the Atmosphere of a ``Hot~Jupiter'' Exoplanet}

\author[ ]{D. V. Bisikalo\footnote{\texttt{bisikalo{@}inasan.ru}}}
\author[ ]{A. A. Cherenkov}

\affil[ ]{Institute of Astronomy, Russian Academy of Sciences, Moscow, Russia}

\date{\vspace{-8ex}}

\newcommand\Lp[1]{$\mathrm{L_#1}$}

\def\gs{$\mathrm{g\,s}^{-1}$}
\def\g{$\mathrm{g}$}
\def\Rpl{\ensuremath{R_{\rm pl}}}
\def\Mpl{\ensuremath{M_{\rm pl}}}
\def\Rj{\ensuremath{R_{\rm Jup}}}
\def\Mj{\ensuremath{M_{\rm Jup}}}

\begin{document}
\maketitle

\begin{abstract}

The results of three-dimensional numerical simulations of the gas dynamics of the atmosphere of a ``hot Jupiter'' exoplanet during the passage of a coronal mass ejection (CME) from the central star are presented. These computations assumed the parameters for the stellar wind and the CME to be typical of the solar values. The characteristic variations of the flow pattern are considered for quasi-closed and closed (but appreciably distorted by the gravitational influence of the star) gaseous envelopes of the exoplanet. It is shown that a typical CME is sufficient to tear off the outer part of an asymmetric envelope that is located beyond the Roche lobe and carry it away from the exoplanet. This leads to a substantial increase in the mass-loss rate from the exoplanet envelope during the passage of CMEs. The mass-loss rate grows by about a factor of 11 for a closed envelope, and by about a factor of 14 for a quasi-closed envelope. Possible evolutionary consequences of the loss of part of the atmosphere during the passage of CMEs are discussed.

\end{abstract}

\section{Introduction}

“Hot Jupiter” exoplanets have masses comparable to the mass of Jupiter and orbital semi-major axes not exceeding 0.1\,AU. The proximity of these objects to their central stars leads to a supersonic regime for the stellar wind flowing around the planet, and therefore to the formation of an outgoing shock followed by a contact discontinuity -- a boundary separating the matter of the wind from the gas of the exoplanetary atmosphere~\citep{Baranov-1977}. According to~\citep{Bisikalo-2013b, Bisikalo-2015}, the gaseous envelopes around hot Jupiters can be divided into three types. First and foremost, we must distinguish closed envelopes, where the head-on collision point (HOCP) corresponding to the minimum distance from the contact discontinuity to the planet is located inside the Roche lobe. Depending on the degree of filling of the Roche lobe, such envelopes may deviate from a spherical shape, but they are undoubtedly characterized by an absence of a significant outflow of matter, $\dot{M} < 10^{9}$\,\gs~\citep{Cherenkov-2014}. If the HOCP, and consequently part of the atmosphere, is located outside the Roche lobe, a significant outflow from the vicinity of the inner Lagrange point \Lp1 is present, and the envelope becomes appreciably asymmetrical. Some parts of the envelope may have restricted dimensions, since the propagation of the flow arising when the Roche lobe overflows may be stopped by the dynamical pressure of the stellar wind. In this case, a quasi-closed, stationary envelope with a complex shape forms in the system~\citep{Bisikalo-2013a}, with corresponding mass-loss rates of $\dot{M} < (3-5) \times 10^{9}$\,\gs~\citep{Cherenkov-2014, Bisikalo-2013c}. If the wind is not able to stop the flow from \Lp1, an open, aspherical envelope forms in the system.

Closed and quasi-closed envelopes are of the most interest for our study, since they are stationary formations with relative low mass-loss rates. Open envelopes are characterized by high mass-loss rates~\citep{Cherenkov-2014} and short lifetimes, and can probably exist only during specific, short-lived stages in the evolution of the atmospheres of hot Jupiters. Previous results on gas-dynamical simulations of the envelopes of hot Jupiters have assumed that the stellar wind does not vary with time. At the same time, it is known from observations that even a relatively quiescent star such as the Sun has a wind in the form of a quasi-neutral plasma that can vary by factors of tens (see, e.g.,~\citep{Zastenker-1999, Liu-2014}).

Beginning in the 1990s, it was known that the main source of the perturbed solar wind is giant ejections of matter from the solar corona, so-called coronal mass ejections (CMEs). According to~\citep{Zastenker-1999, Johnstone-2015, Howard-1985}, CMEs are characterized by a mean mass of plasma ejected into the interplanetary medium $\sim 10^{13}\,\mathrm{kg}$, a mean ejection energy $\sim 10^{31}\,\mathrm{erg}$, and ejection velocities that vary from $100$ to $3000\,\mathrm{km\,s^{-1}}$; i.e., the motion is obviously supersonic, so that the propagation of the ejection is accompanied by the formation of a shock. Moreover, a CME often differs from the ordinary solar wind by an up to 10-15\% increase in the abundance of helium ions. Note that, even for the Sun, the frequency of CMEs is very high, and varies from 0.5 to 2.5 per day, depending on the time separation from the maximum of the solar activity cycle~\citep{Zastenker-1999}. This means that an exopanet located even around a quiescent star will fairly often (no less than twice per month, as is the case for the Earth) be subject to the action of CMEs, which distort the flow patterns in its gaseous envelope.

The main aim of our present study is to investigate the action of CMEs having fairly typical parameters on the structure of flows in the gaseous envelopes of hot Jupiters. The discovery in~\citep{Bisikalo-2013a, Bisikalo-2013b} of the possibility of forming extended gaseous envelopes around hot Jupiters not only fundamentally changes approach to interpreting observational data, but also substantially influences consideration of the physical processes occurring in these envelopes. In particular, it is very obvious that, if the size of the envelope exceeds the size of the Roche lobe, matter that is formally part of the planetary atmosphere will be only weakly gravitationally bound to it. In this case, even a small external action (such as a CME) is sufficient to tear off the outer part of the envelope from the planet, thereby appreciably increasing the mass loss by the exoplanet. Another important consequence of an extended envelope is the possibility of neglecting the influence of the planet's magnetic field on the dynamics of the outer layers of the envelope, and using gas-dynamical equations to describe the flows in the system. Indeed, even with the maximum estimates of the possible magnetic fields on hot Jupiters (up to a tenth of the magnetic moment of Jupiter~\citep{Kislyakova-2014}), the radius of the magnetosphere lies within the Roche lobe, so that the influence of the planet’s magnetic field on the dynamics of outer parts of the envelope is negligible.

We used a modification of the computational code of~\citep{Bisikalo-2013b, Cherenkov-2014}, based on the solution of a system of gas-dynamical equations, for our new computations. We specified the characteristics of the plasma incident on the exoplanet based on the parameters of solar CMEs measured at the Earth’s orbit~\citep{Farrell-2012}. Further, we assumed that the relative variations of the parameters of the CME plasma at the orbit of the hot Jupiter match with the same as at the Earth, and occur at the same moments in time. The absolute values of the parameters were specified assuming that the parameters of the unperturbed wind correspond to the solar values at the orbit of the exoplanet. Computations were carried out for both closed and quasi-closed exoplanet atmospheres. Special attention was given to the variations in the mass-loss rate of the planetary atmosphere under the action of the CME.

The paper is organized as follows. Section~\ref{sec:problem} presents our formulation of the problem. Section~\ref{sec:results} presents the results of our gas-dynamical simulations, and Section~\ref{sec:conclusion} summarizes the main results of our study.

\section{Formulation of the Problem}
\label{sec:problem}

We considered the exoplanet HD\,209458\,b as a typical hot Jupiter. HD\,209458\,b is a transiting exoplanet with radius \Rpl\,=\,1.38\,\Rj\ and mass \Mpl\,= 0.69\,\Mj\ orbiting a main-sequence G0V star at a distance of 0.04747\,AU with an orbital period of 3.52472\,d\footnote{http://exoplanet.eu/}. The parameters of its upper atmosphere (temperature and density) were specified from the range of values determined in~\citep{Koskinen-2013}. We chose two sets of parameters for the computations, corresponding to solutions with closed (but appreciably distorted by the gravitational influence of the star) and quasi-closed atmospheres from~\citep{Bisikalo-2013b}. In the first case, the temperature $T$ was taken to be $7 \times 10^3$\,K and the particle density to be $n = 5 \times 10^{10}\,\mathrm{cm}^{-3}$ at photometric radius; in the second case, $T\,=\,7.5 \times 10^3$\,$\mathrm{K}$ and $n = 10^{11}\,\mathrm{cm}^{-3}$. As in~\citep{Bisikalo-2013b}, the parameters of the stationary wind for the star HD\,209458 were taken to be equal to those of the solar wind at the distance from the star corresponding to the orbit of the planet: $T_w=7.3 \times 10^5\,\mathrm{K}$, $n_w\sim10^4\,\mathrm{cm}^{-3}$, and radial velocity $v_w\,=\,100\,\mathrm{km\,s^{-1}}$~\citep{Withbroe-1988}. With these parameters, the stellar-wind flow is subsonic, with Mach number M\,=\,0.99; however, given the supersonic orbital motion of the planet (M\,=\,1.4), the total velocity of the planet relative to the stellar wind is appreciably supersonic, with Mach number M\,=\,1.75.

The numerical model used to study the flow structure in the envelopes of hot Jupiters is presented in~\citep{Bisikalo-2013b, Cherenkov-2014}. We used a three-dimensional system of equations for the gravitational gas dynamics to describe the envelope, closed by the equation of state of an ideal, neutral, monatonic gas. We assumed that the force field is described by a Roche potential, the rotation of the planet is synchronized with its orbital motion, and the planet has no magnetic field.

We carried out the computations using a Roe--Osher TVD scheme with the entropy correction of Einfeldt (see, e.g.,~\citep{Zhilkin-2012}). The coordinate system used has its origin at the center of the star, the $X$ axis directed along the line joining the centers of the star and planet, the $Z$ axis is perpendicular to the orbital plane, and the $Y$ axis directed to make the coordinate system right-handed. The computations were performed on an exponentially stretched rectangular grid with $468 \times 468 \times 178$ cells in $X$, $Y$, and $Z$, with the size of the computational domain covering $(40 \times 40 \times 10)$\,\Rpl . The mass-loss rate from the planetary envelope was computed as the difference between the fluxes of matter leaving and entering the system (in a parallelepiped specified along the $X$, $Y$, and $Z$ axes and containing the gaseous envelope).

\begin{figure}[t!]
\begin{center}
\centering\epsfig{width=12cm,file=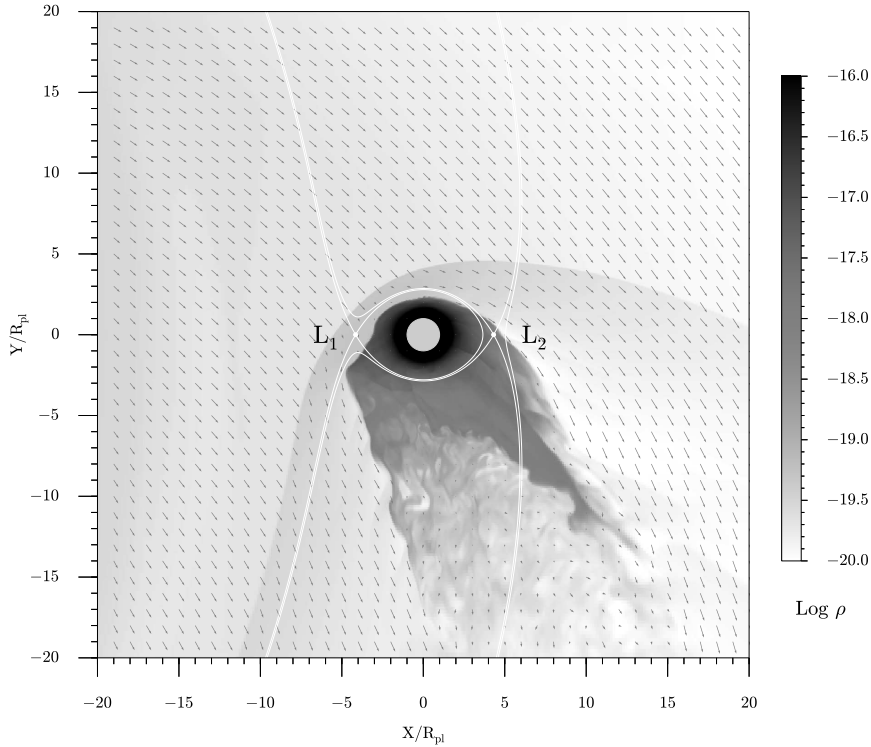}
\end{center}
\caption{Distribution of the density and velocity vectors in a closed exoplanet envelope. A cross section of the envelope in the orbital plane is shown. The star is located to the left. The white circle at the center of the figure represents the planet. The solid white curves denote the contours of the Roche potential, and the positions of the Lagrange points \Lp1 and \Lp2 are indicated.}\label{fig:alone_7000}
\end{figure}

\begin{figure}[t!]
\begin{center}
\centering\epsfig{width=12cm,file=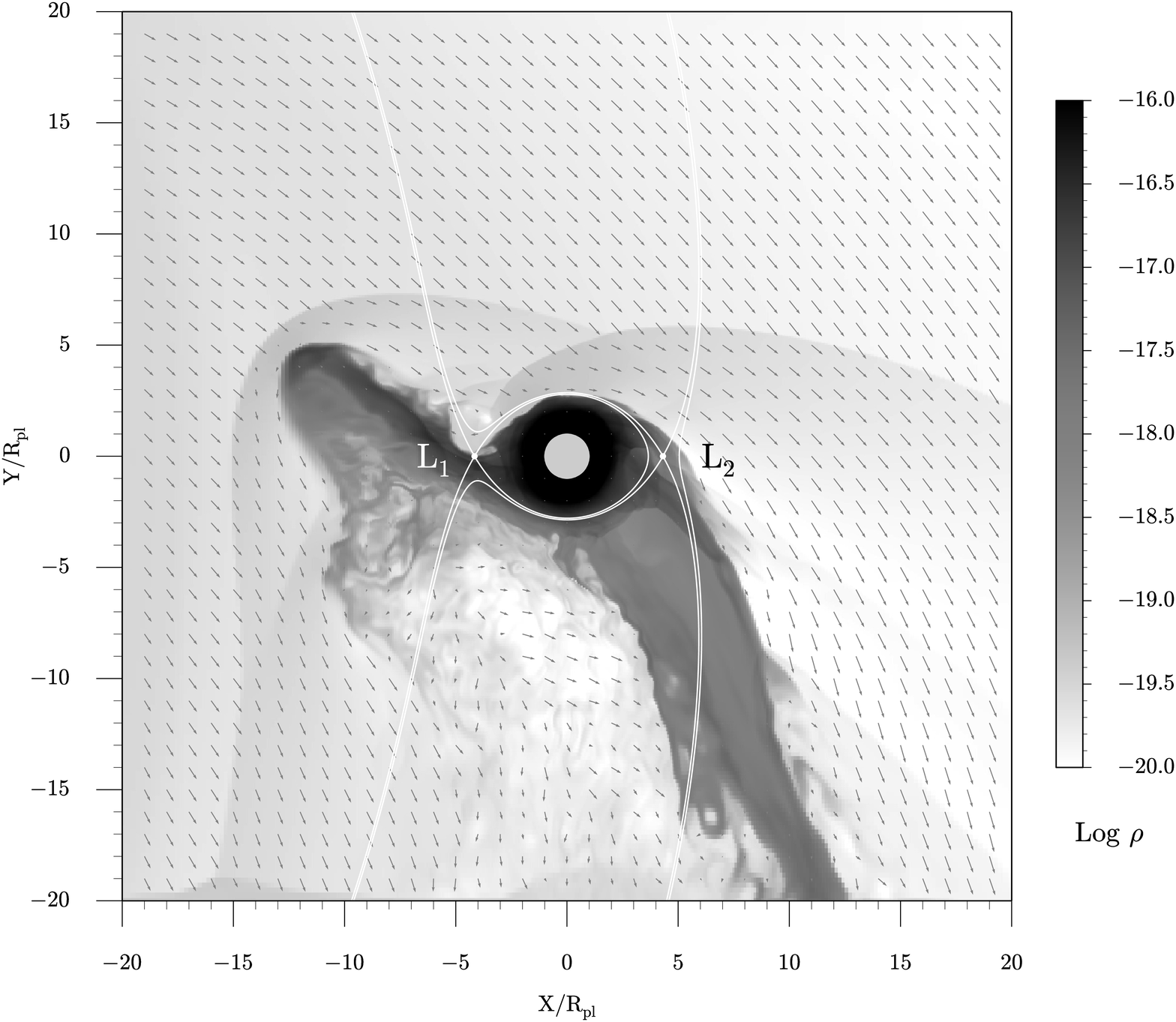}
\caption{Distribution of the density and velocity vectors in a quasi-closed exoplanet envelope. The notation is the same as in Fig.~1.}\label{fig:alone_7500}
\end{center}
\end{figure}

The results of the computations for the case of a stationary stellar wind for closed and quasi-closed envelopes are presented in Figs.~1 and~2. These figures present the distribution of the density and velocity vectors in the orbital plane of the system. Analysis of the overall morphology of the flow indicates that, even for the case of a closed envelope (Fig.~1), with the adopted set of parameters, the atmosphere begins to become aspherical due to the gravitational influence of the star. Two small bulges directed toward the points \Lp1 and \Lp2 form, leading to appreciable distortion of the shape of the shock compared to the solution obtained for an essentially spherical atmosphere (see, e.g.,~\citep{Bisikalo-2013b}). The formation of an envelope that appreciably resembles a Roche lobe shape is observed in the system, however the outflow of matter from the atmosphere remains modest, at a level of $\dot{M} \simeq 2 \times 10^{9}$\,\gs.

Figure~2 depicts the main elements of the flow for the case of a quasi-closed envelope. In this solution, a powerful stream from the vicinity of \Lp1 forms in the direction toward the star, as well as a less intense but significant outflow from the vicinity of \Lp2 away from the star. In this solution, the stream of matter from \Lp1 is stopped by the dynamical pressure of the stellar wind at a distance of several times \Rpl\,\,from the exoplanet. The forming envelope has a complex shape, where, in addition to the atmosphere itself, two bulges associated with the flow toward \Lp1 and \Lp2 can clearly be seen. The supersonic motion of the planet in the gas of the stellar wind leads to the formation of a system of two outgoing shocks. The HOCP of the first shock is located ahead of the flow of matter from \Lp1, and the HOCP of the second shock ahead of the spherical part of the atmosphere. With the adopted parameters for the stationary stellar wind, the total mass-loss rate is approximately $\dot{M} \simeq 3 \times 10^{9}$\,\gs.

To investigate the influence of a CME on the flow structure in the gaseous envelope of a hot Jupiter, we must determine the time variations of the stellar-wind parameters during the passage of the CME and use these as time-dependent boundary conditions. We used measured parameters of the solar wind at the Earth’s orbit obtained with various spacecraft (ACE, WIND, SOHO) during a CME on May 1–4, 1998~\citep{Farrell-2012}. Since it is not possible to obtain measurements at a distance corresponding to the orbit of a hot Jupiter, we assumed that the relative variations of the wind parameters during the CME were the same as those at the Earth’s orbit, and occur at the same times.

\begin{table}[b!]
\caption{Parameters of the stellar wind during the passage
of the CME}\label{tabular:models}
\begin{center}
\begin{tabular}{c|c|c|c|c}
\hline
Phase & 1 & 2 & 3 & 4\\
\hline
\hline
Duration of
the phase
(hours) & - & 8.5 & 13 & 22\\ 
$n \mathrm{ \, (cm^{-3})}$ & $\mathrm{10^4}$ & $\mathrm{4\times10^4}$ & $\mathrm{6\times10^3}$ & $\mathrm{10^5}$ \\
$T \mathrm{ \, (K)}$ & $\mathrm{7.3\times10^5}$ & $\mathrm{3.7\times10^6}$ & $\mathrm{5.8\times10^5}$ & $\mathrm{2.2\times10^5}$ \\
$v \mathrm{ \, (km\cdot s^{-1})}$ & $\mathrm{100}$ & $\mathrm{133}$ & $\mathrm{144}$ & $\mathrm{111}$ \\
\hline
\end{tabular}

\end{center}
\end{table}

According to~\citep{Farrell-2012}, the passage of a CME can be separated into four phases. First, there is the usual stationary solar wind, corresponding to the computations presented in Figs.~1 and~2 for the closed and quasi-closed envelopes. We used these solutions as initial conditions for the next phase. The second phase of the CME begins with the passage of the shock, and is characterized by increases in the plasma density, temperature, and velocity by factors of $\sim\!4$, $\sim\!5$, and $\sim\!1.3$, respectively. The duration of the second phase is 8.5 h. The third phase, or so-called early CME phase, begins with the passage of the contact discontinuity following the shock, and has a duration of 13 h. During this phase, the plasma velocity grows by another factor of $\sim\!1.1$, the temperature falls by a factor of $\sim\!6$, and the density decreases to about $\sim\!0.6$ of its value in the unperturbed wind. The fourth phase, or so-called late-CME phase, does not have a clearly defined beginning, since it is not associated with the passage of a discontinuity. However, the plasma parameters vary substantially in this phase: the density grows by a factor of $\sim\!\!10$ compared to its unperturbed value, providing the basis for distinguishing this stage in the development of the ejection as a distinct phase of the CME and making it possible to mark its onset based on this sharp rise in density. In addition, the fraction of $\mathrm{He}^{++}$ increases from several percent to 20-30\% in this phase of the CME~\citep{Skoug-1999}. The duration of this phase is approximately 22 h. After the end of the fourth phase, the wind parameters return to their initial values.

Applying this division of the CME development into four phases together with measurements of the relative jumps in the solar-wind parameters at the beginning of each phase, we specified the variation of the stellar-wind parameters in the HD\,209458 system to behave in the same way. The main parameters of the wind plasma, specified in the computations as time-dependent boundary conditions, are presented in Table~1.

\bigskip

%


\section{Computational Results}
\label{sec:results}

\begin{wrapfigure}{Lth!}{0.5\textwidth}
\begin{center}
\epsfig{trim={2cm 2cm 0 5cm},clip,width=7cm,file=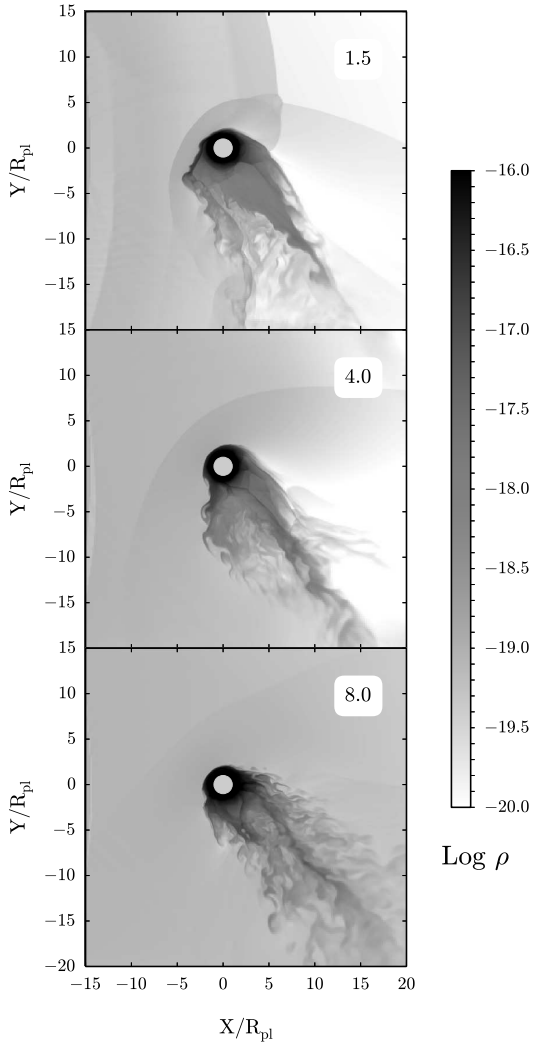}
\end{center}
\caption{Distribution of the density in a closed exoplanet envelope subject to the action of a CME, for various moments in time. As in Fig.~1, a cross section of the envelope in the orbital plane is shown, the star is located to the left, and the white circle at the center represents the planet. The upper, middle and lower diagrams correspond to the beginning, middle and end of the second phase of the CME. The time in hours measured from the onset of the second phase is given in the upper right corner of each diagram.}
\end{wrapfigure}

Figures~3,~4, and~5 show the density distribution in the closed exoplanet envelope subject to the action of a CME for various times covering the active phases of the CME (from the beginning of the second phase to the end of the fourth phase). Each figure presents solutions for three times. As in Fig.~1, each diagram shows a cross section of the envelope in the orbital plane of the system. All the density distributions are shown on the same gray scale, which also coincides with the density scale in Fig.~1. The time for which a solution is shown, measured from the start of the second phase in hours, is shown in the upper right corner of each diagram.

The density distribution depicted in the upper panel of Fig.~3 corresponds to the onset of the second phase of the CME. The second phase has a higher dynamical pressure than the stationary wind in the first phase ($\rho_2 v_2^2 /\rho_1 v_1^2 = 7.1$), and its onset is accompanied by the propagation of a shock across the computational domain. This shock propagates from the star (i.e., from left to right); it is clear from the density distribution that the matter behind the shock front has a high density. During its collision with the gaseous envelope, the shock shifts the vortical wake of the stationary flow. The middle and lower distributions in Fig.~3, corresponding to the middle and end of the second phase, demonstrate that the duration of this phase is sufficient to establish a new quasi-stationary state, but with a higher dynamical pressure of the wind. The increase in the wind velocity leads to a shift in the position of the vortical wake behind the planet, causing it to approach the axis connecting the star and planet and appreciably decreasing its transverse size. It is interesting that, under the action of the denser and faster wind, the asymmetry of the envelope due to the gravitational influence of the star decreases, and the wind interacts with the inner spherical layers of the planetary atmosphere.

The upper, middle, and lower diagrams in Fig.~4 show the density distributions at the beginning, middle, and end of the third phase of the CME. The third phase is characterized by a lower dynamical pressure of the stellar wind, compared to the second phase ($\rho_3 v_3^2 /\rho_2 v_2^2 = 0.2$), leading to a complete change in the structure of the gaseous envelope of the exoplanet. The envelope begins to expand, but does not have time to reach an equilibrium state before the onset of fourth phase of the CME. Note that the characteristic size of the envelope during the third phase appreciably exceeds the size of the closed envelope obtained in the stationary solution.

The fourth phase of the CME has the highest dynamical pressure in the wind ($\rho_4 v_4^2 /\rho_3 v_3^2 = 9.9$, $\rho_4 v_4^2 /\rho_1 v_1^2 = 12.3$). Density distributions corresponding to the beginning, middle, and end of this phase are presented in the upper, middle, and lower diagrams in Fig.~5. Note that our simulations did not take into account variations in the chemical composition of the wind. The high density of the stellar wind leads to the accumulation of matter in the envelope in the initial stage of this phase. The solution arrives at a new quasi-stationary state during the fourth phase. As in the second phase, the higher dynamical pressure of the wind shifts the HOCP closer to the planet, consequently leading to the formation of an envelope whose shape is less distorted by the gravitational influence of the star. After the completion of the fourth phase, the wind parameters were set equal to their initial stationary values, and the solution had returned to the original solution after 25 h.

\begin{figure}[!t]
\centering
\begin{minipage}{.5\textwidth}
  \centering\epsfig{trim={2cm 2cm 0cm 5cm},clip, width=7cm,file=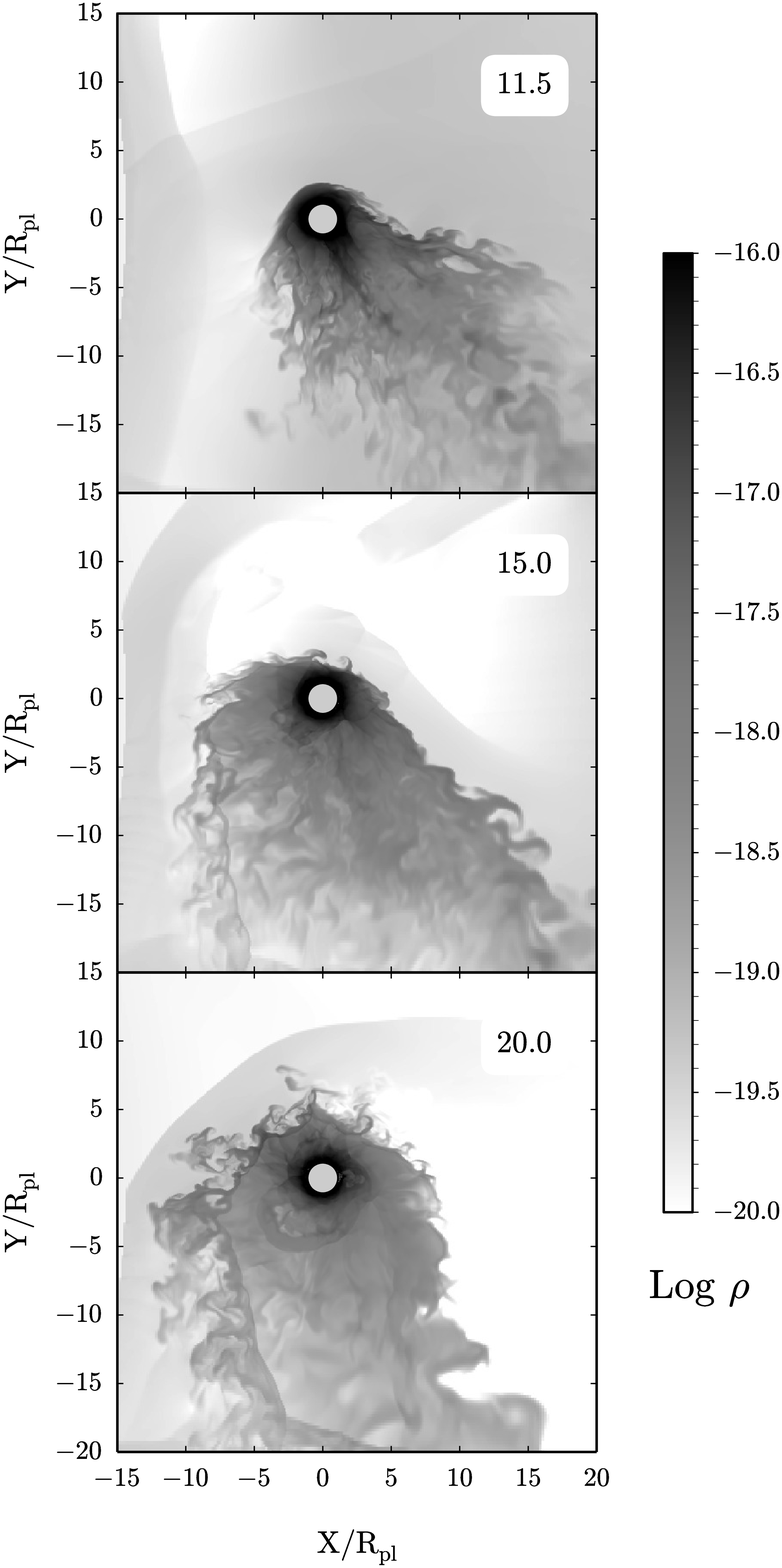}
  \captionsetup{width=0.9\textwidth}
  \captionof{figure}{Same as Fig.~3 for the third phase of the CME.}
\end{minipage}%
\begin{minipage}{.5\textwidth}
  \centering\epsfig{trim={2cm 2cm 0cm 5cm},clip,width=7cm,file=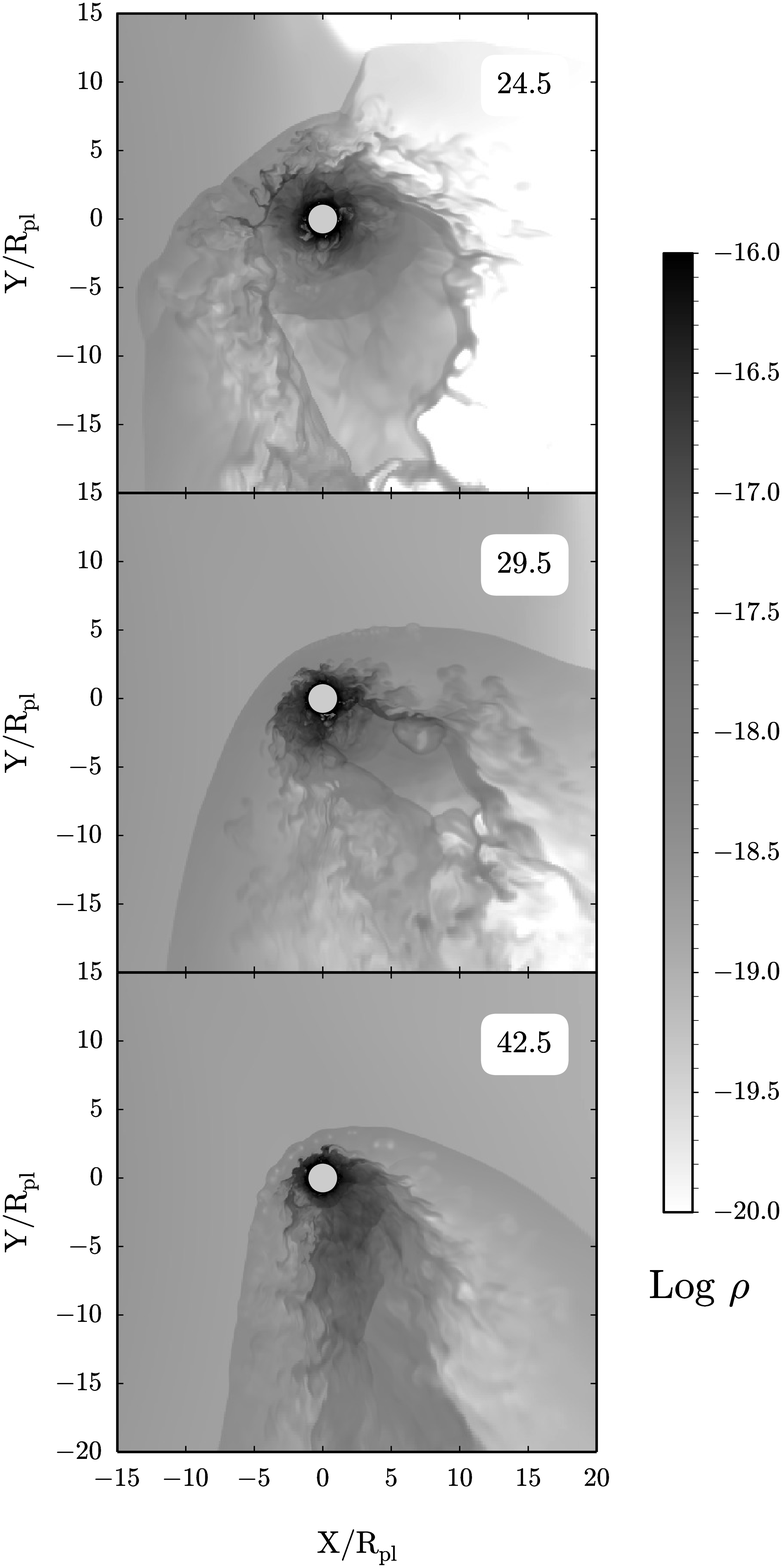}
  \captionsetup{width=0.9\textwidth}
  \captionof{figure}{Same as Fig.~3 for the fourth phase of the CME.}
\end{minipage}
\end{figure}

The mass-loss rate during the passage of the CME around the closed envelope is depicted in Fig.~6. The solid curve shows the computational results. The computations were stopped at a time $t = 0.7 P_{orb}$, since further changes in the solution do not lead to any appreciable variations in the total amount of matter lost by the planet. An extrapolation of the solution from the end of the computations to the emergence into the stationary state is shown by the dot--dashed curve. Apart from short periods at the beginning of the second and fourth phases, when the dense stellar wind arrives at the envelope, the mass-loss rate substantially exceeds its equilibrium value. The mean mass-loss rates for each phase of the CME are presented in Table~2. The last column in Table~2 (labeled~5) corresponds to the return to the stationary state after the end of the passage of the CME. The mass-loss rate obtained in the equilibrium solution is $2 \times 10^{9}$\,\gs, shown in Fig. 6 by the dashed horizontal line in the first and fifth phases. The total amount of matter lost by the exoplanet during the passage of the CME is $5 \times 10^{15}$\,\g. This exceeds the mass lost in the stationary solution during the same time (taking into account the return to the stationary state), $\Delta t \sim 0.77 P_{orb}$, by a factor of 11.

\begin{figure}[t!]
\begin{center}
\centering\epsfig{width=12cm,file=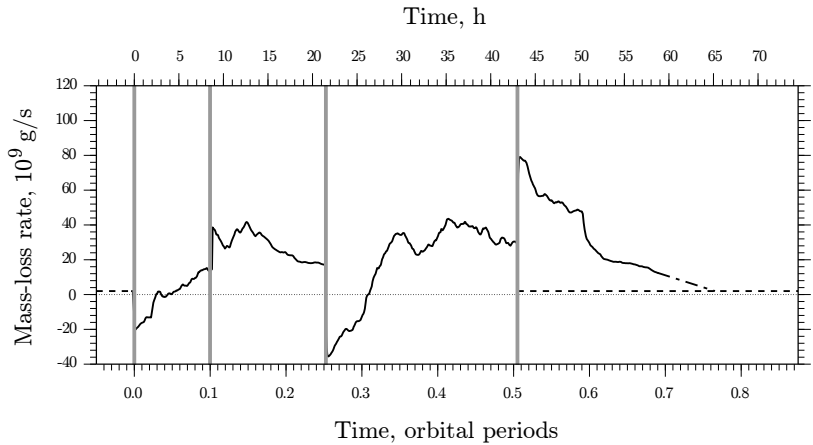}
\end{center}
\caption{Mass-loss rate during the passage of the CME around a closed envelope. The various phases of the CME are indicated
by the vertical gray lines.}\label{fig:flows_7000}
\end{figure}

\begin{figure}[!t]
\centering
\begin{minipage}{.5\textwidth}
  \centering\epsfig{trim={2cm 2cm 0 5cm},clip,width=7cm,file=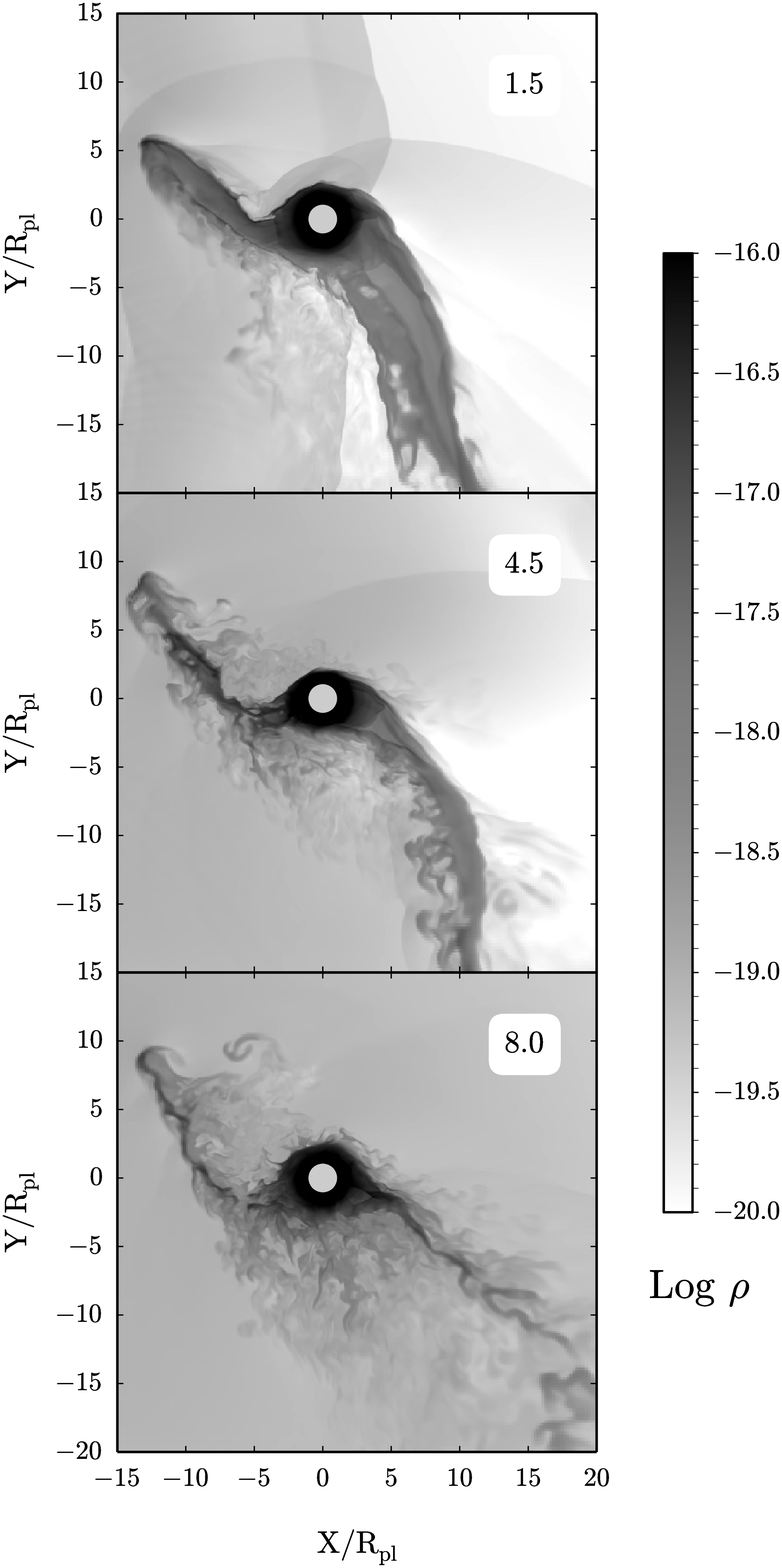}
  \captionsetup{width=0.9\textwidth}
  \captionof{figure}{Distribution of the density in a quasi-closed exoplanet envelope subject to the action of a CME for various moments in time. As in Fig.~2, a cross section of the envelope in the orbital planet is shown, the star is located to the left, and the white circle at the center represents the planet. The upper, middle and lower diagrams correspond to the beginning, middle, and end of the second phase of the CME. The time in hours measured from the onset of the second phase is given in the upper right corner of each diagram.}
\end{minipage}%
\begin{minipage}{.5\textwidth}
  \centering\epsfig{trim={2cm 2cm 0 5cm},clip,width=7cm,file=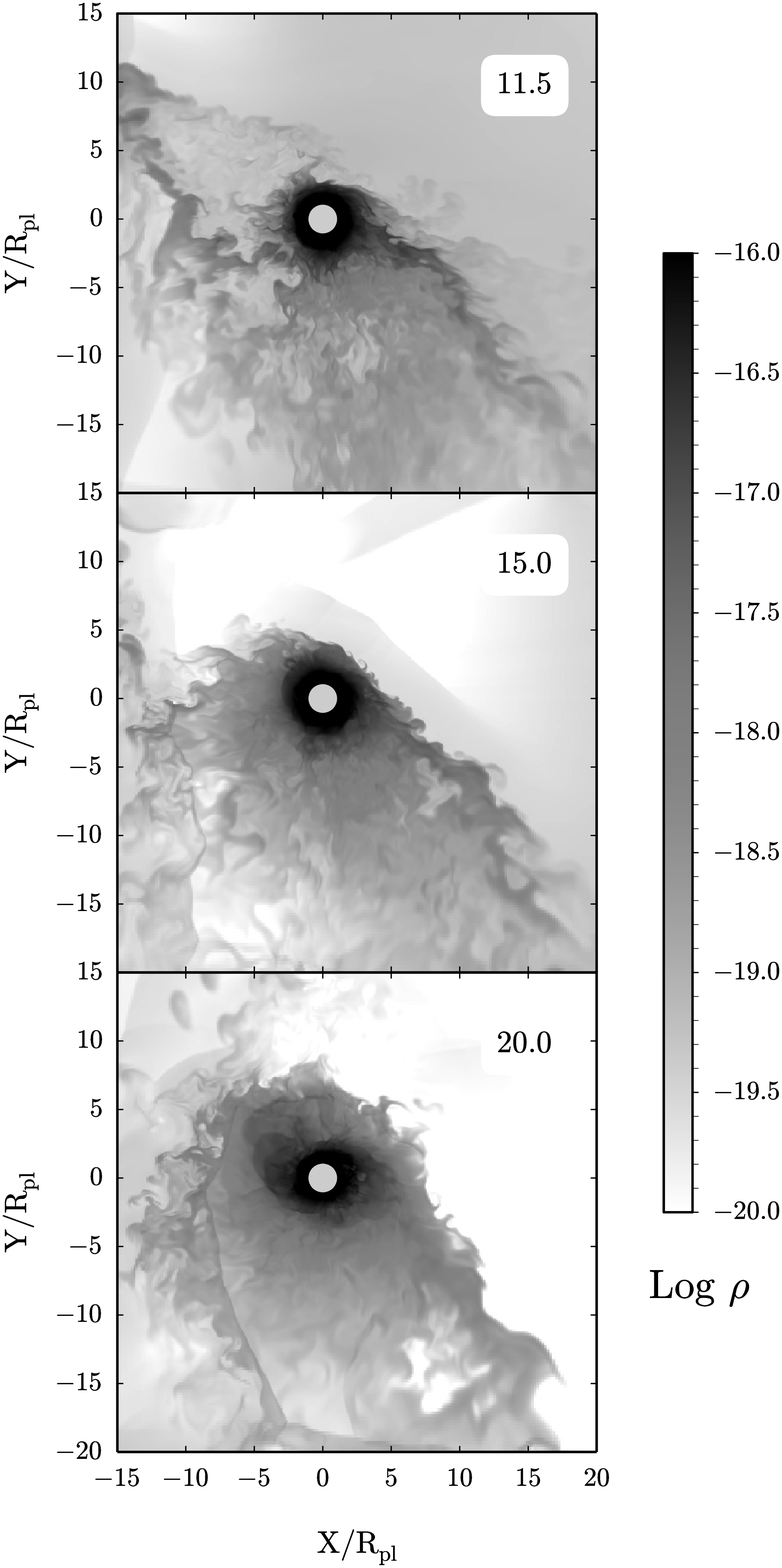}
  \captionsetup{width=0.9\textwidth}
  \captionof{figure}{Same as Fig. 7 for the third phase of the CME.\\ \\ \\ \\ \\ \\ \\ \\ \\ \\}
\end{minipage}
\end{figure}

Figures~7,~8, and~9 show the evolution of the density distribution in a quasi-closed exoplanet envelope subject to the action of a CME. For ease of comparison, all the notation in the figures, coordinates, gray scale, and times for each diagram have been chosen to be the same as in Figs.~3,~4, and~5.

As in the closed atmosphere, the upper diagram in Fig.~7 shows the density distribution corresponding to the onset of the second phase of the CME. This solution is close to the stationary case (the first phase of the CME), and does not differ strongly from the density distribution in Fig.~2. The density distributions shown in the middle and lower diagrams in Fig.~7 demonstrate that, similar to the case of the closed envelope, the dense, fast wind shifts the vortical wake behind the planet and makes it less dense. However, the largest changes occur for the asymmetrical part of the envelope, located ahead of the planet and originating in the vicinity of the inner Lagrange point \Lp1. The wind in the second phase of the CME, which has a high dynamical pressure, begins to disrupt this part of the envelope, tearing small portions off and forming a vortical wake behind this feature in the flow. It is obvious that, if the duration of this phase were longer, this flows feature of the asymmetrical envelope would be disrupted and carried from the system, in spite of the constant feeding of matter from~\Lp1. Note that part of the disrupted stream from~\Lp1 returns to the Roche lobe of the exoplanet, and can be accreted by the atmosphere.

The solutions obtained during the third phase of the CME are presented in the upper, middle, and lower diagrams of Fig.~8, which correspond to the beginning, middle, and end of this phase. Due to the relative decrease in the dynamical pressure of the stellar wind, the remnant of the envelope expands. Moreover, the stream of matter from \Lp1 begins to form an asymmetrical part of the envelope ahead of the planet; however, a new equilibrium state is not reached due to the short duration of this phase.

The density distributions obtained during the fourth phase of the CME are presented in the three diagrams of Fig.~9. A comparison of the solutions corresponding to the beginning, middle, and end of the fourth phase (the upper, middle, and lower digrams in Fig.~9) shows that, as for the closed atmosphere, the solution obtained during this phase arrives at a new quasi-stationary state with a HOCP located closer to the planet. However, in this case, the stream of matter from \Lp1 is sufficiently powerful to hinder the shift of the HOCP toward the Roche lobe. As a consequence, the envelope remains quasi-closed, with clearly visible outflows from the vicinities of \Lp1 and \Lp2, but with appreciably smaller dimensions.

The mass-loss rate during the passage of the CME around a quasi-closed envelope is shown in Fig.~10. As in Fig.~6, the solid curve shows the computational results. An extrapolation of the solution from the end of the simulations ($t = 0.7P_{orb}$) to the emergence to the new stationary state is shown by the dot-dashed curve. As for the closed atmosphere, apart from short periods at the beginning of the second and fourth phases, the mass-loss rate substantially exceeds its equilibrium value during the passage of the CME (see also Table~2). The mass-loss rate obtained in the equilibrium solution is $3 \times 10^{9}$\,\gs, and is shown in Fig.~10 by the dashed line in the first phase. The total amount of matter lost by the exoplanet during the passage of the CME is $1.0 \times 10^{16}$\,\g, which exceeds the mass lost in the stationary solution over the same time (taking into account the emergence onto the stationary state), $\Delta t \sim 0.83 P_{orb}$ , by approximately a factor of 14.


\section{Conclusion}

\label{sec:conclusion}

\begin{wrapfigure}{Lht!}{0.5\textwidth}
\centering
\epsfig{trim={2cm 2cm 0 5cm},clip,width=7cm,file=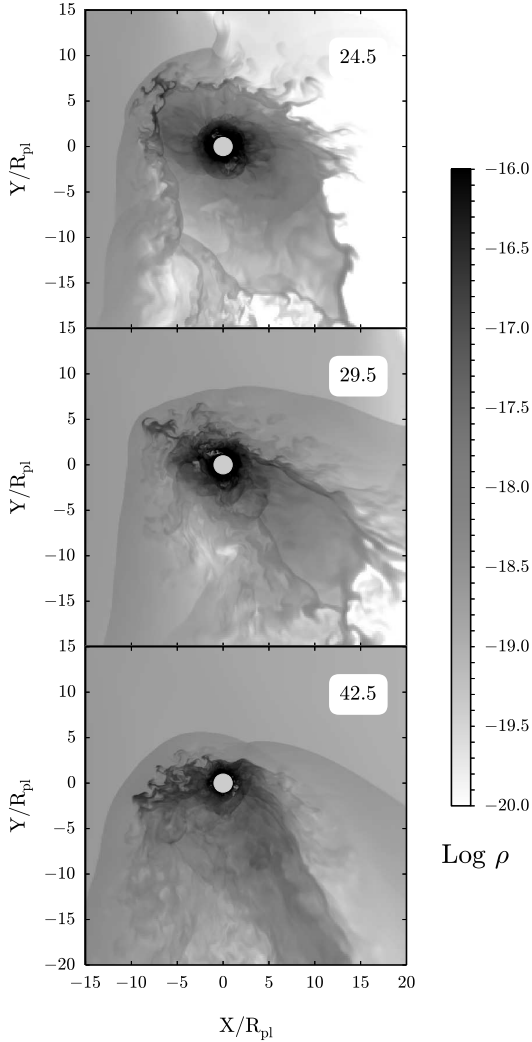}
\caption{Same as Fig.~7 for the fourth phase of the CME.}
\end{wrapfigure}

We have considered the gas dynamics of the interaction of a coronal mass ejection from a star with the atmosphere of a hot Jupiter exoplanet. These computations assumed that the parameters of the stellar wind and the CME corresponded to their solar values. Computations were carried out for both closed (but appreciably distorted by the gravitational influence of the star) and quasi-closed gaseous envelope of an exoplanet. With the adopted parameters for the atmosphere, an appreciable fraction of the envelope is located beyond the Roche lobe in both models. During the interaction with the CME, this part of the envelope, and even some of the envelope that is located within the Roche lobe, becomes gravitationally unbound from the planet, and is ejected from the system.

The results of our three-dimensional numerical simulations have established that the total mass lost during the passage of the CME in the case of a closed envelope is $\Delta M \simeq 5 \times 10^{15}$\,\g. The mass lost in the case of a quasi-closed envelope is $\Delta M \simeq 1.0 \times 10^{16}$\,\g. These masses exceed the masses lost in the stationary solutions for the closed and quasi-closed envelopes over the same time intervals by factors of 10.8 and 14.1, respectively.

Let us consider the possible evolutionary consequences of tearing off part of the envelope of a hot Jupiter during the passage of a CME. Let us suppose that the star displays solar-type activity; i.e., the rate at which CMEs distort the atmosphere of the planet is $\sim\!2$ per month. We will also suppose that the main parameters of the CMEs (the characteristic variations in the density, velocity, temperature, and the durations of the various phases) are equal to those for solar CMEs. In this case, according to our computations, the total mass loss over a year will be increased by factors of 2.7 and 3.5 for the closed and quasi-closed envelopes, compared to their equlibrium values (obtained for a stationary wind). Thus, even for weakly active stars, taking into account the influence of CMEs leads to approximately a threefold decrease in the characteristic lifetime of the atmosphere of a hot Jupiter. Note that the probability of a hot Jupiter having an extended quasi-closed envelope is approximately one-third~\citep{Bisikalo-2015}, so that our estimates should appreciably influence the overall evolution of the atmospheres of such exoplanets. Equally important, the probability and intensity of CMEs increase for young stars (compared to the Sun), further decreasing the lifetimes of extended envelopes around hot Jupiters. Summarizing, the role of CMEs in the evolution of the gaseous envelopes of hot Jupiters is very important, and must be included when considering the evolution of the atmospheres of such exoplanets.

\section*{Acknowledgments}

This work was supported by the Russian Science Foundation (grant RNF 14-12-01048). The results of the work were obtained using computational resources of MCC NRC “Kurchatov Institute” (http://computing.kiae.ru/).

\begin{figure}[]
\begin{center}
\centering\epsfig{width=12cm,file=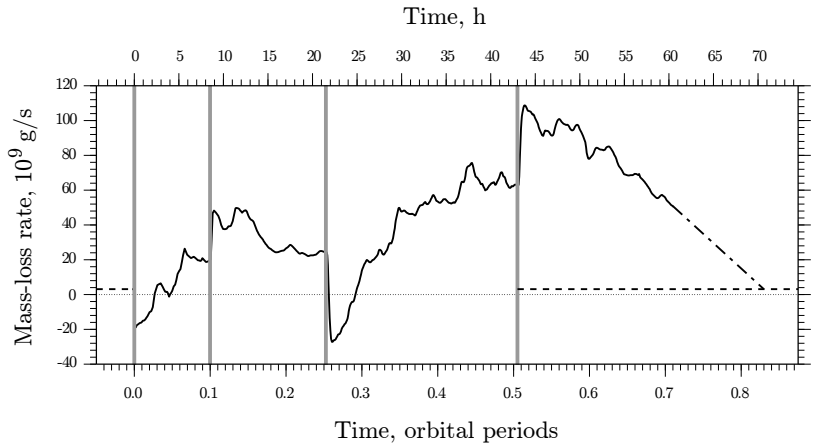}
\end{center}
\caption{Mass-loss rate during the passage of the CME around a quasi-closed envelope. The various phases of the CME are
indicated by the vertical gray lines.}\label{fig:flows_7500}
\end{figure}

\begin{table}[]
\begin{center}
\caption{Mean mass-loss rates for various phases in the passage of the CME (in units of $\times10^9$\,\gs)}\label{tabular:massLoss}
\begin{tabular}{c|c|c|c|c|c}
\hline
Phase & 1 & 2 & 3 & 4 & 5\\
\hline
\hline
Solution for a closed atmosphere & $\mathrm{2.0} $ & $\mathrm{0.0}$ & $\mathrm{27.5}$ & $\mathrm{19.5}$ & $\mathrm{28.3}$\\ 
Solution for a quasi-closed atmosphere & $\mathrm{3.0}$ & $\mathrm{5.9}$ & $\mathrm{33.2}$ & $\mathrm{39.3}$ & $\mathrm{60.1}$\\
\hline
\end{tabular}
\end{center}
\end{table}


\end{document}